\documentclass[prd,amsmath,amssymb,showpacs]{revtex4}
\usepackage[colorlinks=true, linkcolor=red]{hyperref}

\begin{document}
\title{Physics of computation and light sheet concept in the measurement of (4+n)-dimensional spacetime geometry}
\author{Pavle Midodashvili}
\email{pmidodashvili@yahoo.com}
\affiliation{Chavchavadze State
University \\32 Chavchavadze Avenue, Tbilisi 0179, Georgia\\and\\Gori State University\\
53 Chavchavadze St., Gori 1400, Georgia}
\date{\today}
\begin{abstract}
We analyze the limits that quantum mechanics imposes on the accuracy to which $(4+n)$-dimensional spacetime geometry can be measured. Using physics of computation and light sheet concept we derive explicit expressions for quantum fluctuations and explore their cumulative effects for various spacetime foam models.
\end{abstract}

\maketitle


\section{Introduction}
Physicists believe that due to quantum fluctuations spacetime can no longer be smooth on small scales (at lengths of about $10^{ - 35}$ meters), but must be roiling and frothy. On such scales spacetime can be considered as composed of an ever-changing arrangement of bubbles called spacetime foam. Obviously it has influence on the ultimate accuracies of distance and time measurements and on the physics of quantum computation. Because quantum fluctuations of spacetime manifest themselves in the form of uncertainties in the geometry of spacetime, the structure of spacetime foam can be inferred from the accuracy with which we can measure that geometry.

In this article we generalize the work of Ng and collaborators (see \cite{JackNg1,JackNg2} and references therein) to higher dimensions. Considering the question about the accuracy to which one can map out the geometry of spacetimes with $n$ extra spatial dimensions at small and large scales, we follow closely Ng's approach and arguments. In the article we derive explicit expressions for quantum fluctuations and explore their cumulative effects for various spacetime foam models. We also introduce the "uniform model" of spacetime foam, which, in general, is different from random-walk model (these models yield different results in spacetime dimensions different from $(1+3)$). The only ingredients used in the discussion are quantum mechanics and general relativity, the two pillars of modern physics. In particular, to derive uncertainties in the mapping of large-scale structure of spacetime we use the physics of computation and light sheet concept. We hope that the results are very general and have wide validity.
\section{Black holes, Planck units in $(4+n)$-dimensional spacetime}
To begin with we recall some known facts for black holes in $(4+n)$-dimensional spacetime, \textit{i.e.} in spacetime with $n$ extra spatial coordinates. The Schwarzschild solution in $(4+n)$ dimensional spacetime has the form \cite{Myers-Perry}
\begin{equation}\label{Schwarzschild}
\begin{array}{l}
ds^2  =  - \Phi \left( r \right)c^2 dt^2  + \Phi (r)^{ - 1} dr^2  + r^2 d\Omega _{2 + n}^2 , \,\,\,\,\,\Phi (r) = 1 - \left( {r_{ \rm{S}} /r} \right)^{1 + n} , \\  r_{\rm{S}}  = \left( {B \frac{{G m}}{{c^2 }}} \right)^{\frac{1} {{1 + n}}} , \,\,\,\,\, B  = \frac{{16\pi }}{{(2 + n)A }}, \,\,\,\,\,A  = \frac{{2\pi ^{\frac{3 + n}{2}}}}{{\Gamma \left( {\frac{3 + n}{2}} \right)}}, \\
\end{array}
\end{equation}
where $r_{\rm{S}}$ and $G$ are Schwarzschild radius, associated with the mass m, and gravitational constant in $(4+n)$-dimensional spacetime respectively. Here $A$ denotes the area of the unit $(2+n)$-sphere in $(3+n)$-dimensional space and $\Gamma \left( x \right) = \int_0^\infty  {e^{ - t} t^{x - 1}dt }$ is the gamma function. The $(4+n)$-dimensional spacetime Planck length, time and mass are defined as follows:
\begin{equation}\label{PlanckUnits}
l_{\rm{P}}  = \left( {\frac{{\hbar G }}{{c^3 }}} \right)^{\frac{1}{{2 + n}}} ,\,\,\,\,\,t_{\rm{P}}  = \frac{{l_{\rm{P}} }}{c},\,\,\,\,\,m_{\rm{P}}  = \frac{\hbar }{{cl_{\rm{P}} }}\,.
\end{equation}
 For $n=0$ these formulae give well known values in ordinary $4$-dimensional spacetime.

 \section{Minimum distance and time that can be measured in higher dimensional spacetimes}
 Let us consider the limits on the minimum distance and time that can be measured. In the case of time measurement we can use the Margolus-Levitin theorem \cite{Mar-Lev} in quantum computation. Indeed, any clock used in time measurement can be considered as a quantum system of some energy $E$ that is available for its functioning. According to the Margolus-Levitin theorem the time $\delta t$ it takes the quantum system to go from one state  to an orthogonal state is greater or equal to $\frac{{\pi \hbar }}{E}$. So the minimum "tick length" of a clock with energy $E$ is $\delta t = \frac{{\pi \hbar }}{E}$, and the precision of clock used in time measurement can be increased by increasing its energy. As regards the distance measurement the wavelength of the particle used to map out space also can be decreased by increasing their energy. This process can be continued until one reaches the Planck scale, see eq.~(\ref{PlanckUnits}), when the Compton wavelength $\frac{{2\pi \hbar }}{{mc}}$ of the clock or particle used in measurements gets the same order of magnitude as their Schwarzschild radius $r_{\rm{S}}  = \left( {B \frac{{G m}}{{c^2 }}} \right)^{\frac{1} {{1 + n}}}$, see eq.~(\ref{Schwarzschild}). That is the realm  of quantum gravitational effects. So the Planck length and the Planck time give natural cutoff at small length and time scales.

\section{Mapping out the large-scale geometry of higher dimensional spacetimes}
In this section we consider the question about the accuracy to which one can map out the large-scale structure of higher dimensional spacetime. Here we generalize the work of Ng and collaborators (see \cite{JackNg1,JackNg2} and references therein) to higher dimensions, and in considering the question we follow closely Ng's approach and arguments. We will use the physics of computation and light sheet concept to derive the limits. Because quantum fluctuations of spacetime manifest themselves in the form of uncertainties in the geometry of spacetime, the structure of spacetime foam can be inferred from the accuracy with which we can measure that geometry.

Let us consider a gedanken process of mapping out the geometry of spacetime for a spherical $(3+n)$-volume of radius $l$ over the amount of time $t$. One way to do this is to fill the space with some distributed measuring system (for example, distributed system of clocks (see \cite{JackNg2})), exchanging signals between its parts and measuring the signal's times of arrival. Such distributed system can be considered as a computer that maps out the geometry of spacetime by means of computation, in which distances and time are gauged by transmitting and processing information.

The total number of elementary operations that the measuring system can perform per second is limited by the Margolus-Levitin theorem \cite{Mar-Lev}, namely the rate of operations for any computer cannot exceed the amount of energy $E$ that is available for computation divided by $\pi \hbar /2$. As the presence of Planck's reduced constant suggests, this rate limit is fundamentally quantum mechanical, and is actually attained by quantum computers \cite{Lloyd}. So, if $M$ denotes the total mass of our distributed measuring system involved in a process of mapping the geometry of spacetime then, via the Margolus-Levitin theorem, the bound on the total number of elementary operations performed in a $(3+n)$-volume of radius $l$ over the amount of time $t$ is given by $\frac{{2tMc^2 }}{{\pi \hbar }}$. But to prevent black hole formation, $M$ must be less than $\frac{1}{{B}}\frac{{c^2 l^{1 + n} }}{{G}}$. Together, these two limits imply that the total number of elementary operations $N_{\rm{ops}}$ that can occur in a spatial volume of radius $l$ during a time period $t$ is no greater than $\frac{2}{{\pi B}}\frac{{l^{1 + n} ct}}{{l_{\rm{P}}^{2 + n} }}$, \textit{i.e.}
\begin{equation}\label{MaxNumbOfOps}
N_{{\rm{ops}}}  \le \frac{{l^{1 + n} ct}}{{l_{\rm{P}}^{2 + n} }}\,,
\end{equation}
where and henceforth we drop multiplicative factors of order $1$.

The elementary operations can be regarded as partitioning the spacetime volume into elementary "cells" (which can be associated with bits of information). Obviously there can be various partitionings. In the case of partitioning with maximal spatial resolution (which is known as the "holographic model" \cite{JackNg2}) each elementary operation is associated with distinct spatial cell, \textit{i.e.} each spatial cell flips only once during the total measurement time period $t$, and on the average each cell occupies a spatial volume no less than
\begin{equation}\label{MinVolumeOfCells}
\delta V_{\rm{h.m.}} = \frac{{V_{l} }}{{N_{\rm{ops}} }} \ge \frac{{l^2 l_{\rm{P}}^{2 + n} }}{{ct}}\,,
\end{equation}
where $V_{l}=\frac{\pi^{\frac{3+n}{2}}l^{3+n}}{\Gamma \left ( 1+\frac{3+n}{2} \right )} \sim l^{3+n}$ is the total volume of a spherical $(3+n)$-dimensional spatial region of radius $l$. The average spatial separation between the cells can be interpreted as the average minimum uncertainty in the measurement of a distance $l$, that is,
\begin{equation}\label{UncertaintyInDistance}
\delta l_{\rm{h.m.}} \ge  \left( {\frac{{l^2 l_{\rm{P}}^{2 + n} }}{{ct}}} \right)^{\frac{1}{{3 + n}}} .
\end{equation}
As it follows from eq.~(\ref{UncertaintyInDistance}) in the holographic model of spacetime foam the average minimum uncertainty $\delta l_{\rm{h.m.}}$ can be decreased at will taking amount of total measurement time $t$ large enough. This vagueness can be resolved by the concept of the light sheets \cite{Bousso}, that defines the  realistic value of $t$ depending on a spherical $(3+n)$-volume under mapping.

As it is known in a $(4+n)$-dimensional spacetime codimension $2$ spatial surface $S$ of a codimension $1$ spatial region (independently of the shape and location of $S$) possesses precisely four orthogonal null directions: \textit{future directed ingoing}, \textit{future directed outgoing}, \textit{past directed ingoing}, and \textit{past directed outgoing}. These directions can be represented by null hypersurfaces that border on $S$ and are always uniquely generated by the past and future directed light rays orthogonal to $S$. At least two of the four null hypersurfaces can be selected as light sheets, according to the condition of nonpositive expansion \cite{Bousso}.

Let us discuss in detail how light sheets are constructed in our case, \textit{i.e.} for a spherical $(3+n)$-volume of radius $l$ (spacelike codimension $1$ hypersurface) with surface $S$ (codimension $2$ hypersurface). One must keep in mind that our codimension $2$ spatial surface $S$ denotes a surface at some instance of time, for example at $t=0$. The future directed light rays coming from the surface $S$ towards the center of the volume generate a first confined null hypersurface $H_{1}$. This surface can be physically described as follows. Imagine that the surface $S$ is filled with light bulbs that all simultaneously flash up at $t=0$. As the light rays travel towards the center of the volume they generate $H_{1}$. Obviously light rays can be also send towards the past, but we might prefer to think of these as arriving from the past, \textit{i.e.}, a light bulb in the center of the spherical $(3+n)$-volume flashed at an appropriate time for its rays to reach the surface $S$ at $t=0$. This light rays orthogonal to $S$ will generate the second confined null hypersurface $H_{2}$. This two confined null hypersurfaces are light sheets of our spherical $(3+n)$-volume.

The light sheet comprises the corresponding matter system completely, in the same sense in which a $t=const$ surface comprises the system completely. So a light sheet represents another way of taking a snapshot of matter system - in light cone time. Actually it comes much closer to how the system is really observed in practice.

So in eqs.~(\ref{MaxNumbOfOps}) and (\ref{UncertaintyInDistance}) we must take $t=l/c$ corresponding to light sheet formation time interval, \textit{i.e.} the time it takes light to cross the distance from center of spherical volume to its surface.(It must be mentioned that the same physics was used implicitly in the \cite{Lloyd-Ng}.) Then we get

\begin{equation}\label{MaxNumbOfOpsCorrect}
N_{{\rm{ops}}}  \le \frac{{l^{2 + n} }}{{l_{\rm{P}}^{2 + n} }}\,,
\end{equation}

\begin{equation}\label{UncertaintyInDistanceCorrect}
\delta l_{{\rm{h.m.}}}  \ge  \left( {ll_{\rm{P}}^{2 + n} } \right)^{\frac{1}{{3 + n}}}.
\end{equation}
As we mentioned above this case corresponds to the well known holographic model of spacetime foam \cite{JackNg3,Pavle1,Pavle2}, when in a spherical spatial region of size $l$ we have maximum number of spatial cells allowed by holographic principle in $(4+n)$-dimensional spacetime, \textit{i.e.} the number of spatial cells is proportional to $l^{2+n}/l_{\rm{P}}^{2+n}$ (see eq.~(\ref{MaxNumbOfOpsCorrect})).

It is interesting to consider another case of partitioning (we call it "uniform model") that corresponds to spreading the spacetime cells uniformly in both space and time, \textit{i.e.}
\begin{equation}\label{LinkForUniformCase}
\delta l_{{\rm{u.m.}}}=c \delta t_{{\rm{u.m.}}},
\end{equation}
where $\delta l_{{\rm{u.m.}}}$ is the spatial size of a cell and $\delta t_{{\rm{u.m.}}}$ is the time between its two consecutive flips during the measurement. In this case number of spatial cells and the average volume of each cell are as follows
\begin{equation}\label{UniformCells}
N_{{\rm{u.m.cells}}}  = \frac{{N_{\rm{ops}} }}{{\left( {\frac{l}{{c\delta t_{{\rm{u.m.}}}}}} \right)}} \le \frac{{c\delta t_{{\rm{u.m.}}}l^{1 + n} }}{{l_{\rm{P}}^{2 + n} }}\,,\end{equation}
\begin{equation}\label{MinVolOfBits}
\delta V_{\rm{u.m.}} = \left( {\delta l_{\rm{u.m.}}} \right)^{3 + n}  = \frac{{V_l }}{{N_{{\rm{u.m.cells}}} }} \ge \frac{{l^2 l_{\rm{P}}^{2 + n} }}{{c\delta t_{{\rm{u.m.}}}}}\,,
\end{equation}
and, taking into account eq.~(\ref{LinkForUniformCase}), for the number of spatial cells, average spatial volume of each cell and uncertainty in distance measurement we find
\begin{equation}\label{UniformBits}
N_{{\rm{u.m.cells}}}  \le \left( {\frac{l}{{l_{\rm{P}} }}} \right)^{\frac{{\left( {2 + n} \right)\left( {3 + n} \right)}}{{4 + n}}}
\end{equation}

\begin{equation}\label{AverVolOfEachBit}
\delta V_{\rm{u.m.}} \ge \left( {l^2 l_{\rm{P}}^{2 + n} } \right)^{\frac{{3 + n}}{{4 + n}}}\,,
\end{equation}
\begin{equation}\label{UniformCaseDistanceUncertainty}
\delta l_{\rm{u.m.}} \ge  \left( {l^2 l_{\rm{P}}^{2 + n} } \right)^{\frac{1}{{4 + n}}}\,.
\end{equation}
As it is seen from eq.~(\ref{LinkForUniformCase}), in the uniform model of spacetime foam each spatial cell flips once in the time it takes to communicate with a neighboring cell. It is also interesting to find the average number of flips $\overline n _{\rm{u.m.flips}}$ of each spatial cell during the total time of measurement. Obviously $\overline n _{\rm{u.m.flips}}  = \frac{{N_{\rm{ops}} }}{{N_{\rm{u.m.cells}} }}$, and, taking into account eqs.~(\ref{MaxNumbOfOpsCorrect}) and (\ref{UniformBits}), we get $\overline n _{\rm{u.m.flips}} \sim \left( {\frac{l}{{l_P }}} \right)^{\frac{2+n}{4+n}}$. So in the uniform model of spacetime foam each spatial cell on average flips $\left( {\frac{l}{{l_P }}} \right)^{\frac{2+n}{4+n}}$ times during the total time of measurement.

There are many other models of spacetime foam. In general they can be parameterized according to $\delta l_{\alpha} \sim l^{1 - \alpha } l_{\rm{P}}^\alpha$ with $\alpha$ of order $1$ (see \cite{JackNg2}). For the holographic and uniform models this parameter is equal to $\alpha_{\rm{h.m.}}=\frac{2+n}{3+n}$ and $\alpha_{\rm{u.m.}}=\frac{2+n}{4+n}$ respectively.

For $\alpha=\frac{1}{2}$ the uncertainty in the measurement of a distance $l$ is proportional to $l^{\frac{1}{2}}$. Such dependence is characteristic for a random-walk fluctuations, and corresponding model of spacetime foam is known as the random-walk model \cite{JackNg1,JackNg2}. So for the random-walk model we have
\begin{equation}\label{RandomWalkCaseDistanceUncertainty}
\delta l_{\rm{r.w.m.}} \sim l^{\frac{1}{2}} l_{\rm{P}}^{\frac{1}{2}},
\end{equation}
and, as it is seen from eq.~(\ref{UniformCaseDistanceUncertainty}), uniform model and random-walk model are identical only in the case $n=0$ (\textit{i.e.} in the ordinary $4$-dimensional spacetime) in complete accordance with \cite{JackNg2}. In general the uniform model of spacetime foam is different from the random-walk model of spacetime foam, and these models yield different results in spacetime dimensions different from $(1+3)$. In the case of random-walk model the number of spatial cells contained in a spherical spatial region of size $l$ is equal to $N_{\rm{r.w.m.cells}}\sim \frac{l^{3+n}}{\delta l_{\rm{r.w.m.}}^{3+n}}=\left( {\frac{l}{{l_P }}} \right)^{\frac{{3 + n}}{2}}$, and for the average number of flips of each cell during the total measurement time $t=\frac{l}{c}$ we have
\begin{equation}\label{FlipsInRWMDurTotTime}
\overline n _{\rm{r.w.m.flips}}=\frac{{N_{\rm{ops}} }}{{N_{\rm{u.m.cells}} }} \sim \left( {\frac{l}{{l_P }}} \right)^{\frac{{1 + n}}{2}},
\end{equation}
where $N_{\rm{ops}}$ is defined in eq.~(\ref{MaxNumbOfOpsCorrect}). On the other hand, if $\delta t_{\rm{r.w.m.}}$ denotes the average time period between two successive flips of a spatial cell in the random-walk model, then
\begin{equation}\label{Connection}
\overline n _{\rm{r.w.m.flips}}=\frac{t}{\delta t_{\rm{r.w.m.}}}
\end{equation}
and, using $t=\frac{l}{c}$ and eq.~(\ref{FlipsInRWMDurTotTime}), it's easy to find from eq.~(\ref{Connection})
\begin{equation}\label{TimePeriod}
\delta t_{\rm{r.w.m.}}\sim \frac{l}{c}\left( {\frac{l}{{l_P }}} \right)^{-\frac{{1 + n}}{2}}.
\end{equation}
Now from eqs.~(\ref{RandomWalkCaseDistanceUncertainty}) and (\ref{TimePeriod}) we have
\begin{equation}\label{FlipsInRandomWalkDuringNeighbouringComm}
\frac{{\delta l_{\rm{r.w.m.}} }}{{c\delta t_{\rm{r.w.m.}}}} \sim \left( {\frac{l}{{l_{\rm{P}} }}} \right)^{\frac{n}{2}},
\end{equation}
\textit{i.e.} in the case of random-walk model each spatial cell flips $\left( {\frac{l}{{l_{\rm{P}} }}} \right)^{\frac{n}{2}}$ times in the time it takes to communicate with a neighboring cell.

In a computation performed during the process of mapping out the geometry of a spacetime region the number of operations is always greater than or equal to the number of spatial cells used in the computation. This follows from the fact that for a cell to be used in computation, it has to participate in at least one operation. So the total number of spatial cells is maximal in the holographic model (because in this case the total number of operations that can occur in a spatial region is equal to the total number of spatial cells), and the parameter $\alpha$ of any model must obey inequality $\alpha \le \alpha_{\rm{h.m.}}=\frac{2+n}{3+n}$.

\section{Cumulative effects of spacetime fluctuations}
In this section we consider the cumulative effects of spacetime fluctuations over a large distance, to be exact we generalize the results of \cite{JackNg2} to the higher dimensional spacetimes and follow closely Ng's approach. Take a distance $l$, and divide it into $n=\frac{l}{\lambda}$ equal parts each of which has length $\lambda \ge l_{\rm{P}}$. The question is how do the $n$ uncertainties $\delta \lambda$ from each part add up to total uncertainty $\delta l$ for the whole distance. The effect is characterized by cumulative factor $C$ defined by $\delta l=C\delta \lambda$. Since for any fixed spacetime foam model $\delta l \sim l^{1 - \alpha } l_{\rm{P}}^\alpha$ and $\delta \lambda \sim \lambda^{1 - \alpha } l_{\rm{P}}^\alpha$, the corresponding cumulative factor is $C_{\alpha}=\left(\frac{l}{\lambda}\right)^{1-\alpha}$. For the random-walk model, when the individual fluctuations $\delta \lambda$ (which take $\pm$ sign with equal probability) from $n=\frac{l}{\lambda}$ parts in $l$ are completely random, the cumulative factor is $C_{\rm{r.w.m.}}=\left(\frac{l}{\lambda}\right)^{\frac{1}{2}}$.

For the holographic case successive individual fluctuations appear to be entangled and somewhat anti-correlated. In this case a plus fluctuation is slightly more likely followed by a minus fluctuation and vice versa, so that together individual fluctuations produce a total fluctuation less than that in the random-walk model. The cumulative factor is $C_{\rm{h.m.}}=\left(\frac{l}{\lambda}\right)^{\frac{1}{3+n}}$.

For the uniform model the cumulative factor is $C_{\rm{u.m.}}=\left(\frac{l}{\lambda}\right)^{\frac{2}{4+n}}$, and successive individual fluctuations are entangled only in spacetimes with $n\ne0$ (in the case of $4$-dimensional spacetime (\textit{i.e.} $n=0$) the uniform model entirely coincides with the random-walk model).

When $n\ne0$ these cumulative factors obey the inequality $C_{\rm{h.m.}}<C_{\rm{u.m.}}<C_{\rm{r.w.m.}}$, \textit{i.e.} the maximal spatial resolution in the mapping of spacetime geometry is achieved for the holographic model.

Finally note, that if $n \to \infty$ then $C_{\rm{h.m.}} \to 1$ and $C_{\rm{u.m.}} \to 1$, \textit{i.e.} in this limiting case successive fluctuations (for the holographic and uniform models) are completely anti-correlated, and the total fluctuation in the measurement of any distance $l$ does not depend on $l$ and is equal to the Planck length $l_{\rm{P}}$.


\end{document}